\documentclass{IJCAS}
\usepackage{url}
\usepackage{array,tabularx}
\usepackage{multicol} 
\usepackage{graphicx}      
\usepackage[numbers]{natbib}       
\usepackage{amsmath}
\usepackage{amssymb, mathrsfs}
\usepackage{comment}

\journalvolumn{VV}
\journalnumber{X}
\journalyear{YYYY}
\setarticlestartpagenumber{1}

\allowdisplaybreaks

\begin{document}
\title{Data-Enabled Predictive Control with Predictive Adaptive Line-of-Sight Guidance for 3-D Path Following of Autonomous Underwater Vehicles}

\author{Sebastian Zieglmeier*, Mathias Hudoba de Badyn, Narada Warakagoda, Thomas Krogstad, and Paal Engelstad}

\begin{abstract}
This paper presents a fully data-driven 3-D path-following framework for autonomous underwater vehicles (AUVs), a representative class of underwater field robotics, based on Data-Enabled Predictive Control (DeePC). The approach eliminates explicit hydrodynamic modeling by exploiting measured input-output trajectories to predict and optimize future system behavior. Classic DeePC is employed for heading control, while a cascaded DeePC architecture with loop-frequency separation is proposed for depth regulation, extending DeePC to plants whose dominant output evolves significantly slower than the actuator bandwidth. For 3-D waypoint path following, the Adaptive Line-of-Sight (ALOS) guidance law is extended to a predictive multistep formulation (PALOS) that supplies the horizon-consistent reference required by receding-horizon predictive controllers. All methods are validated in high-fidelity 6 degrees of freedom simulation on the REMUS~100 AUV under nominal operation, ocean-current disturbances, operation beyond the data regime, and 3-D waypoint path following, consistently outperforming the corresponding state-of-the-art benchmarks. In 3-D waypoint path following, the framework reduces cross-track error by approximately 28\% relative to the ALOS-PI/PID baseline.
\end{abstract}


\maketitle

\makeAuthorInformation{
Sebastian Zieglmeier, Mathias Hudoba de Badyn and Paal Engelstad are with the Department of Technology Systems, University of Oslo, 2027 Kjeller, Norway (e-mail: sebastiz@uio.no, mathihud@uio.no). Narada Warakagoda and Thomas Krogstad are with the Norwegian Defence Research Establishment, 2027 Kjeller, Norway.

* Corresponding author.
}

\runningtitle{2026}{Sebastian Zieglmeier et. al.}{Data-Enabled Predictive Control and Guidance for Autonomous Underwater Vehicles}{xxx}{xxxx}{x}

\section{Introduction}
As a representative class of underwater field robotics, Autonomous Underwater Vehicles (AUVs) have become central to modern marine exploration, inspection, and defense activities~\citep{Gu_2022}. Ensuring reliable path-following and trajectory tracking in such missions requires controllers that can cope with nonlinear hydrodynamics and environmental disturbances~\citep{Shen_2020}. These control objectives are typically achieved through a hierarchical control structure, which integrates a high-level guidance system responsible for generating appropriate reference signals, an intermediate-level controller tracking those references, and a low-level control system that directly manages the actuators~\citep{Breivik_2009}.
A classic example of a high-level guidance system for AUVs is Line-of-Sight (LOS) guidance. The LOS method provides a geometric three-point guidance rule for steering the vehicle toward a target point on the desired path, effectively translating the spatial error into a heading command~\citep{Breivik_2009, Borhaug_2006}. More sophisticated LOS-based guidance laws have been summarized in~\citep{Gu_2022}, and the state-of-the-art lies in the Adaptive Line-of-Sight (ALOS) algorithm, which optimizes reference generation by compensating for varying environmental conditions~\citep{Fossen_2023}.
In practice, these high-level guidance laws are commonly paired with reactive PI/PID controllers that do not account for future behavior or constraints. As a result, their performance deteriorates when faced with strong nonlinearities, actuator saturation, or delays in dynamics. Consequently, predictive control strategies have been introduced to AUVs, such as Model Predictive Control (MPC)~\citep{Shen_2020, Rout_2020}.

These approaches require a model of the AUV dynamics, which is complex and time-consuming~\citep{Taubert_2014}. The hydrodynamic coefficients that characterize added mass, drag, lift, and damping are often nonlinear, coupled, and highly dependent on factors such as vehicle geometry, speed, and surrounding flow conditions. Identifying these parameters requires extensive experiments that are time-consuming and context-specific~\citep{Taubert_2014}, and the resulting models typically remain valid only within a limited operating range, reducing their applicability across different speeds, payloads, or environmental conditions.
Such challenges have motivated data-driven control approaches, which exploit measured data to design and execute control laws without requiring an explicit parametric model of the system. Within this context, the Data-Enabled Predictive Control (DeePC) framework introduced in~\citep{Coulson_2019} has emerged as one of the most promising approaches. DeePC uses offline trajectories of inputs and outputs to solve a constrained optimization problem for optimal control inputs. This is particularly attractive for AUV applications, especially due to its potential to also handle complex and nonlinear dynamics~\citep{Elokda_2021}. 

This work develops a fully data-driven 3-D path-following framework for AUVs, with the following control-methodological contributions:
\begin{itemize}
    \item \textbf{Cascaded DeePC architecture with loop-frequency separation:} For AUV missions with externally prescribed depth and heading references, direct application of DeePC to the depth response is impractical. Within feasible data lengths, the slow heave dynamics cannot be sufficiently excited to satisfy the persistency-of-excitation requirement. A cascaded data-driven architecture is proposed that decouples depth regulation into an outer DeePC mapping the desired heave position to a pitch reference and an inner DeePC regulating pitch via the stern-plane deflection, with the two loops operating at separated frequencies. A two-stage data-collection procedure together with a unity-gain assumption for the inner loop enables the persistency of excitation required by the outer DeePC, extending the applicability of DeePC to plants whose dominant output channel evolves significantly slower than the actuator bandwidth.
    \item \textbf{Predictive adaptive line-of-sight guidance (PALOS):} For waypoint-based 3-D path-following missions, the single-step output of the classical ALOS guidance law is incompatible with the horizon-length reference required by receding-horizon predictive controllers such as DeePC. A predictive extension is proposed that recursively forward-propagates the vehicle position using the commanded attitudes as the predicted orientation, evaluates the ALOS law at each predicted position, and updates the active waypoint pair whenever transitions occur within the prediction horizon. The result is a horizon-consistent guidance reference that anticipates path transitions and integrates directly into the DeePC optimization.
    \item \textbf{Validation of integrated data-driven 3-D path-following framework:} The proposed control architectures are validated through a sensitivity analysis in a high-fidelity 6-DOF simulation environment, spanning nominal operation, ocean-current disturbances, operation beyond the data regime, and 3-D waypoint path following, consistently outperforming the corresponding state-of-the-art benchmarks.
\end{itemize}
The Marine Systems Simulator (MSS) implementation of the REMUS~100~\citep{fossen2004mss, fossen2021handbook} used in this work captures the dominant nonlinear hydrodynamic phenomena acting on the vehicle: added-mass effects, Coriolis-centripetal coupling, velocity-dependent damping, lift and drag from the control surfaces, cross-flow drag along the hull, attitude-dependent restoring forces, and ocean-current disturbances. As an established reference platform in the AUV control literature, this digital-twin-grade representation provides sufficient complexity for establishing the architectural viability of DeePC prior to hardware deployment, requiring costly operational logistics.\\
This paper is organized as follows.
Notation is outlined in Section~\ref{sec:notation}, and the general AUV dynamics, state-of-the-art, and theoretical foundations of DeePC are in Section~\ref{sec:auv}-\ref{sec:deepc}, respectively. 
Subsequently, the control architecture design and PALOS are described in Section~\ref{sec:deepc_auv}-\ref{sec:palos}, and numerical results, including stress test, obtained from high fidelity MSS-based simulations are analyzed in Section~\ref{sec:results}. The built-in PID controllers and the recently developed ALOS algorithm~\citep{Fossen_2023, Fossen_2024} from the MSS toolbox are employed as benchmarks to contextualize the obtained results. Finally, the conclusion and future works are in Section~\ref{sec:conc}.

\section{Notation}
\label{sec:notation}
$x_k \in \mathbb{R}^n$ denotes the value of $x$ at the discrete time step $k$, where $\mathbb{R}$ denotes the set of real numbers and $n$ the dimension. 
The operator $\operatorname{diag}\{x\}$ denotes a diagonal matrix whose diagonal entries are given by the elements of the vector $x$.\\ 
$\| x\|_M^2$ denotes the squared weighted euclidean norm of the vector $x$ with the positive weighting matrix $M$, where $\| x \|_M^2 := x^\top Mx$.
The stacked column vector of a sequence $\{x\}_a^b$ is denoted as $x := \begin{bmatrix} x_a^\top  ... x_b^\top  \end{bmatrix} ^\top $. \\
The position coordinates of the AUV in the North-East-Down (NED) frame are denoted by $x_p, y_p, z_p$, where the subscript $p$ is introduced to avoid confusion with the conventional control-system state $x$ and output $y$.

\section{Autonomous Underwater Vehicle}
\label{sec:auv}
The AUV considered in this study is the REMUS~100, as implemented in Fossen's MSS toolbox~\citep{fossen2004mss}. The model includes six degrees of freedom (6-DOF) dynamics with actuation provided by a three-bladed propeller, stern, and rudder fins, allowing control of surge, heave, and sway motions. The REMUS~100 is a representative small vehicle with a weight of approximately 31.9~kg, a length of 1.6~m, and a diameter of 0.19~m. 
The AUV dynamics follow the modeling framework of~\citep{fossen2021handbook}, with the specific REMUS~100 parameters obtained from the MSS toolbox~\citep{fossen2004mss}. The dominant relations are outlined in the following, while the corresponding hydrodynamic derivations are reproduced from~\citep{fossen2021handbook} in Appendices~\ref{app:intro}--\ref{app:ocean_current_model}. For the complete formulation, the reader is referred to the original handbook.

\subsection{The rigid-body equations of an AUV}
The dynamics of the AUV are represented using a standard 6-DOF rigid-body formulation~\citep{fossen2021handbook}. The vehicle state $x=[\eta^\top , \nu^\top ]^\top $  is defined by its pose $\eta$ and velocity vector $\nu$. The pose consists of its position $x_p,y_p,z_p$ in the inertial NED frame and the attitude, consisting of the Euler angles $\phi, \theta, \psi$, which describe the orientation of the body frame relative to the NED frame in~\eqref{eq:eta}. The linear velocities $u$, $v$, $w$ (surge, sway, heave) and angular velocities $p$, $q$, $r$ (roll, pitch, yaw) in the velocity vector $\nu$ are expressed in the body frame given in~\eqref{eq:nu}. The relative velocity with respect to the surrounding water is given in~\eqref{eq:relative}, where $\nu_c = [u_c, v_c, w_c, 0, 0, 0]^\top $ contains the ocean currents expressed in the body frame.
\begin{subequations}\label{eq:state_definitions}
\begin{align}
\eta &= [x_p,y_p,z_p, \phi, \theta, \psi]^\top \in \mathbb{R}^6\label{eq:eta} \\
\nu &= [u, v, w, p, q, r]^\top \in \mathbb{R}^6 \label{eq:nu} \\
\nu_r &= \nu - \nu_c \label{eq:relative}
\end{align}
\end{subequations}
The kinematic relation in~\eqref{eq:kinematic} describes the vehicle motion in the inertial NED frame and depends on the absolute body-fixed velocity $\nu$. 
In contrast, the dynamic model in~\eqref{eq:hydrodynamics} is expressed in terms of the relative velocity $\nu_r$, since the hydrodynamic forces depend on the motion of the vehicle relative to the surrounding water. 
\begin{subequations}\label{eq:motion}
\begin{gather}
\dot{\eta} = J(\eta)\nu \label{eq:kinematic} \\
M\dot{\nu}_r + C(\nu_r)\nu_r + D(\nu_r)\nu_r + g(\eta) = \tau + \tau_{LD} + \tau_{CF} \label{eq:hydrodynamics}
\end{gather}
\end{subequations}
\subsection{AUV dynamic matrices}
The detailed derivation of the dynamic components describing the AUV is provided in Appendices~\ref{app:intro} and~\ref{app:dynamics}.
The kinematic equation, given in~\eqref{eq:kinematic}, is defined by the transformation matrix $J(\eta)$, mapping body-fixed linear and angular velocities to the time derivatives of the vehicle’s position and orientation in the NED frame.
The inertia matrix $M$ of the AUV is composed of the rigid-body inertia $M_{RB}$, depending on the vehicle’s mass properties, and the added-mass matrix $M_A$, depending on the hydrodynamic contribution caused by the acceleration of the surrounding fluid. 
In the MSS toolbox, $M_A$ is assumed constant, resulting in a constant $M$ as in~\eqref{eq:Inertia_matrix}.
\begin{equation}
\scalebox{0.8}{$
{M}=
\begin{bmatrix}
m_{11} & 0 & 0 & 0 & m_{15} & 0 \\
0 & m_{22} & 0 & m_{24} & 0 & 0 \\
0 & 0 & m_{33} & 0 & 0 & 0 \\
0 & m_{42} & 0 & m_{44} & 0 & 0 \\
m_{51} & 0 & 0 & 0 & m_{55} & 0 \\
0 & 0 & 0 & 0 & 0 & m_{66}
\end{bmatrix}
$}
\label{eq:Inertia_matrix}
\end{equation}
The Coriolis--centripetal matrix $C(\nu_r)$ in~\eqref{eq:Cori_sum} represents gyroscopic and cross-coupling effects arising from linear and angular motion. It is composed of the rigid-body term $C_{RB}(\nu_r)$, depending on the angular velocities in $\nu_r$,  and the added-mass term $C_A(\nu_r)$, depending on the relative linear velocities in $\nu_r$. Both dependencies contribute to the nonlinear character of the AUV dynamics. 
\begin{equation}
C(\nu_r) = C_{RB}(\nu_r) + C_A(\nu_r)
\label{eq:Cori_sum}
\end{equation}
The damping matrix $D(\nu_r)$ is modeled as a diagonal matrix, representing damping in each degree of freedom with the damping coefficients $X_u, Y_v, Z_w, K_p, M_q, N_r$. To account for nonlinear effects at higher speeds, the damping in the surge and sway directions decreases in the REMUS~100 simulation exponentially with the total relative velocity $U_r$, introducing velocity-dependent behavior in the damping matrix. The resulting formulation is  
\begin{equation}
\scalebox{0.85}{$
D(\nu_r)=-\operatorname{diag}\Big\{\,X_u e^{-3U_r},\;Y_v e^{-3U_r},\;Z_w,\;K_p,\;M_q,\;N_r\,\Big\},
$}
\label{eq:Damping}
\end{equation}
where the damping coefficients $X_u, \dots, N_r$ are derived from the inertia entries and characteristic damping time constants $T_i$, as detailed in Appendix~\ref{app:dynamics}.
The vector $g(\eta)$ represents the restoring forces and moments acting on the vehicle due to gravity and buoyancy, resulting in nonlinear attitude-dependent behavior. For neutrally buoyant underwater vehicles such as the REMUS~100, these effects are primarily determined by the position of the center of gravity, defined as $r_{bG} = [x_g, y_g, z_g]^T$ and the gravitational force $m g$, relative to the body-fixed origin as in~\eqref{eq:submerge}.
\begin{equation}
\scalebox{0.85}{$
{g}({\eta})=\left[\begin{array}{c}
\mathbf{0}_{3\times 1}\\
z_{\mathrm{g}} mg \cos (\theta) \sin (\phi)-y_{\mathrm{g}} mg \cos (\theta) \cos (\phi) \\
z_{\mathrm{g}} mg \sin (\theta)+x_{\mathrm{g}} mg \cos (\theta) \cos (\phi) \\
x_{\mathrm{g}} mg \cos (\theta) \sin (\phi)-y_{\mathrm{g}} mg \sin (\theta)
\end{array}\right]
$}
\label{eq:submerge}
\end{equation}
\subsection{AUV Forces and Moments}
The control input vector is defined as $u = [\delta_s, \delta_r, n_p]^\top$, where $\delta_s$ and $\delta_r$ denote the stern and rudder plane deflections, and $n_p$ denotes the propeller speed. These inputs generate the generalized forces and moments $\tau$ acting on the AUV. In particular, $\tau$ results from the lift and drag forces produced by the stern and rudder planes, i.e., $X_s, X_r, Y_r, Z_s$, together with the propeller thrust $X_{\mathrm{prop}}$ and torque $K_{\mathrm{prop}}$, explicitly stated in Appendix~\ref{app:dynamics}.
The resulting generalized force vector is given in~\eqref{eq:tau}, where $x_r$, $x_s$ are the longitudinal positions of the rudder and stern planes relative to the body-fixed origin. 
\begin{equation}
\tau = \left[ X_{\mathrm{prop}} + X_r + X_s, 
Y_r , 
Z_s , 
K_{\mathrm{prop}}, 
- x_s Z_s, 
x_r Y_r \right]^{\!\top}
\label{eq:tau}
\end{equation}
The lift--drag vector $\tau_{LD}$ in~\eqref{eq:tau_LD} consists of the longitudinal and vertical components $\tau_{LD,X}$ and $\tau_{LD,Z}$, which represent the combined lift and drag forces expressed in the body-fixed frame.
\begin{equation}
\tau_{LD}=\left[
\tau_{LD,X}, 0, \tau_{LD,Z}, 0,0,0 
\right]^\top
\label{eq:tau_LD}
\end{equation}
The total generalized cross-flow drag vector acting on the vehicle is given in~\eqref{eq:tau_CF}, representing the hydrodynamic forces and moments in sway, heave, pitch, and yaw due to lateral and vertical flow separation along the hull.
\begin{equation}
\tau_{CF}= \left[0, \, Y_{CF}, \, Z_{CF}, \, 0, \, M_{CF}, \, N_{CF}\right]^\top
\label{eq:tau_CF}
\end{equation}

\subsection{Ocean current dynamics}
\label{sec:ocean_current_model}
Ocean currents are modeled as irrotational, such that only the linear velocity components contribute to the relative motion between the AUV and the surrounding water. The current in the NED frame is decomposed into a horizontal component with magnitude $V_c$ and direction $\beta_{V_c}$, and a vertical component $W_c$, given in~\eqref{eq:current_simplified_NED_short}.
Further in-depth modeling details are provided in Appendix~\ref{app:ocean_current_model}.
\begin{equation}
{\nu}_c^n = 
\begin{bmatrix}
V_c \cos(\beta_{V_c}),\;
V_c \sin(\beta_{V_c}),\;
W_c
\end{bmatrix}^\top.
\label{eq:current_simplified_NED_short}
\end{equation}

\section{Classical PI/PID AUV Control}
\label{sec:classic}
The classical PI/PID architecture, the operational standard for AUV tracking control~\citep{Gu_2022, fossen2021handbook}, serves as the tracking-layer benchmark in this work. Parametric Model Predictive Control would in principle provide a closer tracking comparison in predictive capability, but requires the hydrodynamic identification effort~\citep{Taubert_2014} that the data-driven framework is designed to eliminate, so its implementation would itself constitute a separate development effort rather than an established benchmark.  The controller gains below were tuned by pole placement as in~\citep{fossen2021handbook, Beard_2012}, with the propeller speed $n_p$ fixed at a nominal cruising velocity for path-following.
\begin{figure}[ht]
\begin{center}
\includegraphics[width=\linewidth]{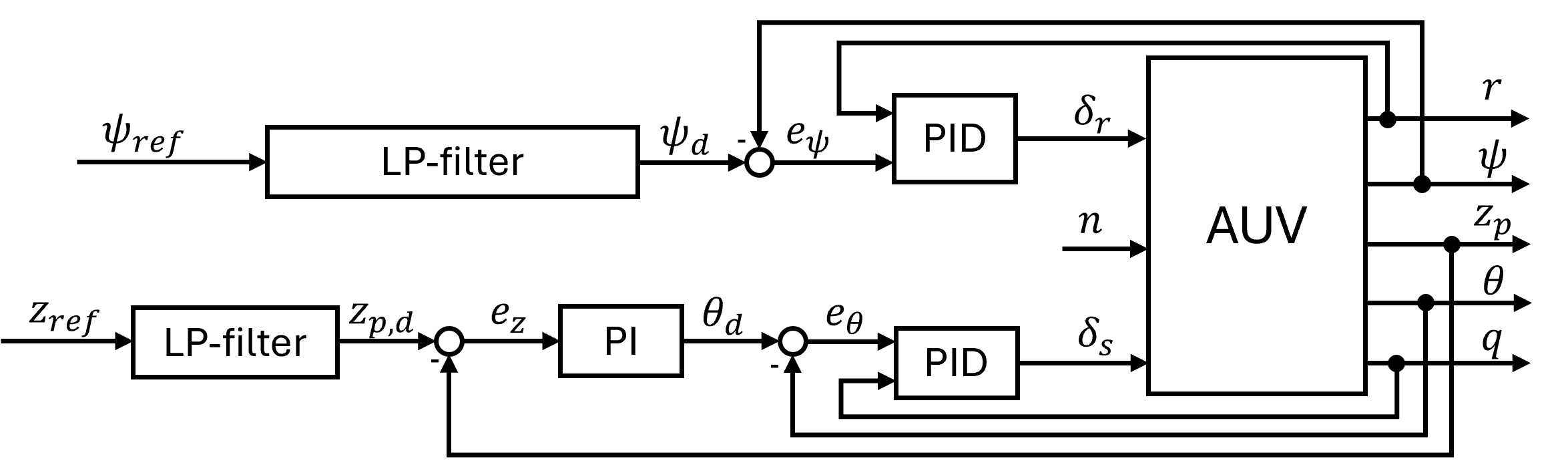}    
\caption{Classic AUV control architecture with the PID heading control and the cascaded PI/PID control of the heave position with an inner loop controlling the pitch angle.} 
\label{fig:control_structure}
\end{center}
\end{figure}
\subsection{Heading Control}
The heading control loop (upper loop in Fig.~\ref{fig:control_structure}) employs a low-pass (LP) filter to generate a smooth desired yaw angle~$\psi_d$ from the commanded heading~$\psi_{\mathrm{ref}}$. 
Using in addition the measured yaw angle~$\psi$ and yaw rate~$r$, the rudder input~$\delta_r$ is computed by a PID controller according to~\eqref{eq:heading_PI}.
\begin{equation}
\delta_r = -K_{p_\psi}(\psi - \psi_d) - K_{d_\psi} r 
           - K_{i_\psi} { \int_0^t (\psi - \psi_d)\, \mathrm{d}\tau}
\label{eq:heading_PI}
\end{equation}
\subsection{Depth Control}
The depth control loop (lower loop in Fig.~\ref{fig:control_structure}) employs a cascaded structure.  
An LP filter smooths the reference $z_{\mathrm{ref}}$ to a desired heave position $z_{p,d}$.  
The outer PI controller generates a desired pitch angle $\theta_d$ from the depth error $e_z=z_p-z_{p,d}$, while the inner PID controller regulates $\theta$ with the pitch rate $q$ via the stern-plane deflection $\delta_s$:
\begin{equation}
\begin{aligned}
\theta_d &= K_{p_z}(z_p-z_{p,d}) + K_{i_z}{\int_0^t (z_p-z_{p,d})\,\mathrm{d}\tau}\\
\delta_s &= -K_{p_\theta}(\theta-\theta_d) - K_{d_\theta}q 
            - K_{i_\theta}{\int_0^t(\theta-\theta_d)\,\mathrm{d}\tau}
\end{aligned}
\label{eq:PI_depth}
\end{equation}
\section{Data-Enabled Predictive Control}
\label{sec:deepc}
The DeePC framework, introduced in~\citep{Coulson_2019}, relies on measured input–output trajectories that are sufficiently informative to capture the system dynamics, a property known as persistency of excitation~\citep{Willems_2004}.  
According to Willems’ Fundamental Lemma, any trajectory of a controllable and observable LTI system can be expressed as a linear combination of previously recorded input–output data.  
The collected input and output sequences, $u_{\mathrm{data}}$ and $y_{\mathrm{data}}$, are arranged into block Hankel matrices $\mathscr{H}(u)$ and $\mathscr{H}(y)$ with $r$ block rows and $c$ columns, as defined in~\citep{Coulson_2019}. Alternatively, the data can be structured as a Page matrix $\mathscr{H}_P(u)$ as in ~\eqref{eq:mosaic-Hankelmatrix}, which groups consecutive columns into separate pages~\citep{Coulson_2021}.
\begin{equation}
\scalebox{0.83}{$
\mathscr{H}(u)\!=\!\left[\!
\begin{array}{@{}c@{}c@{}c@{}c@{}}
u_1 & u_2 & \cdots & u_c \\
\vdots & \vdots & \cdots & \vdots \\
u_r & \,u_{r+1} & \,\cdots & u_{r+c-1}
\end{array}
\!\right],
\mathscr{H}_P(u)\!=\!\left[\!
\begin{array}{@{}c@{}c@{}c@{}c@{}}
u_1 & \,u_{r+1} & \,\cdots & u_{(c-1)r+1} \\
\vdots & \vdots & \cdots & \vdots \\
u_r & u_{2r} & \cdots & u_{cr}
\end{array}
\!\right]
\label{eq:mosaic-Hankelmatrix}
$}
\end{equation}
General persistency of excitation requires these data matrices to be of full row rank~\citep{Coulson_2019}. For noisy data, a quantitative characterization of this condition was introduced in~\citep{Coulson_2022_1}, relating the degree of excitation to the singular value of the data matrix.  
For the DeePC formulation, the data matrices are partitioned into past and future components as in~\eqref{eq:UdYd}, where $(U_p, Y_p)$ encode past trajectories for initialization, and $(U_f, Y_f)$ define the future trajectories used for prediction.
\begin{equation}
U_d = 
\begin{bmatrix} U_p \\ U_f \end{bmatrix},
\qquad
Y_d = 
\begin{bmatrix} Y_p \\ Y_f \end{bmatrix}
\label{eq:UdYd}
\end{equation}
Consider a discrete-time system with inputs $u \in \mathbb{R}^{m}$ and outputs $y \in \mathbb{R}^{p}$ over a prediction horizon $N$.  
Let $u_{\mathrm{ini}}$ and $y_{\mathrm{ini}}$ denote the past input–output trajectories of length $T_{\mathrm{ini}}$, and $u$ and $y$ the future control inputs and predicted outputs, respectively.  
The DeePC framework formulates the following finite-horizon optimization problem:
\begin{equation}
\scalebox{0.9}{$
\begin{array}{ll}
\underset{g, u, y, \sigma}{\operatorname{min}} 
&  \|r - y \|_Q^2 + \| u \|_R^2 
+  \| g \|_{\lambda_g}^2 +  \| \sigma_y \|_{\lambda_y}^2 \\
\text {s.t.} &
\left(\begin{array}{c}
U_{{p}} \\ Y_{{p}} \\ U_{{f}} \\ Y_{{f}}
\end{array}\right) g
=
\left(\begin{array}{c}
u_{{\mathrm{ini}}} \\ y_{{\mathrm{ini}}} \\ u \\ y
\end{array}\right)
+
\left(\begin{array}{c}
0 \\ \sigma_y \\ 0 \\ 0
\end{array}\right), \\[8pt]
& u_k \in \mathcal{U}, \; y_k \in \mathcal{Y}, \quad \forall k \in \{0, \ldots, N-1\}.
\end{array}
$}
\label{eq:DeePC}
\end{equation}
Here, $g \in \mathbb{R}^{T - T_{\mathrm{ini}} - N + 1}$ is the decision vector that linearly combines past trajectories to reconstruct the system behavior, and $r \in \mathbb{R}^{p}$ denotes the reference trajectory.  
$Q \in \mathbb{R}^{p \times p}$ and $R \in \mathbb{R}^{m \times m}$ are positive definite weighting matrices.  
The regularization terms $\textstyle{\| g \|_{\lambda_g}^2 }$ and $\textstyle{  \| \sigma_y \|_{\lambda_y}^2}$, with $\lambda_g, \lambda_y > 0$, follow the approach of~\citep{Huang_2023} to improve robustness against measurement noise and nonlinearities. 
Input and output constraints are directly incorporated through the admissible sets $\mathcal{U}$ and $\mathcal{Y}$.
Theoretical and experimental guarantees regarding robustness and stability are provided in \citep{Berberich_2020, Berberich_2021, Berberich_2024}. 

\section{DeePC for AUV}
\label{sec:deepc_auv}
\begin{figure}[ht]
\begin{center}
\includegraphics[width=\linewidth]{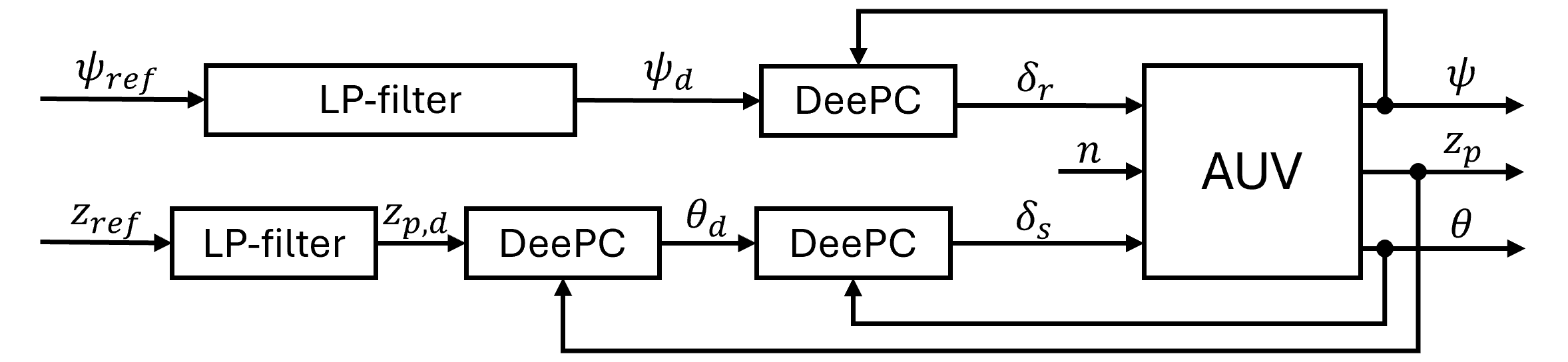}    
\caption{Control architecture of the DeePC-based heading control and the cascaded DeePC control of the heave position with an inner loop controlling the pitch angle.} 
\label{fig:block_diagram}
\end{center}
\end{figure}

This section develops the data-driven control architectures for AUV missions with externally prescribed depth and heading references. A fully coupled DeePC formulation incorporating all three control inputs and twelve states could, in principle, capture the complete 6-DOF dynamics of the AUV, provided that sufficiently rich data persistently excite all relevant dynamic modes. However, the dynamics of the different states operate in different frequency regimes, and exhibit distinct damping behavior, which makes a fully coupled formulation unsuitable for AUV control. Consequently, this work adopts a decoupled control architecture, consistent with classical AUV control principles, comprising independent heading and depth control loops, as illustrated in Fig.~\ref{fig:block_diagram}. The data-driven formulation of DeePC accommodates the nonlinear coupling between the 6-DOF channels implicitly through the recorded input-output sequences, sidestepping the construction of explicit nonlinear models that impede parametric MPC approaches on AUVs.
\subsection{DeePC Heading Control}
For a direct performance comparison, the DeePC-based heading control employs the same LP reference filter as the classical PID structure. To generate a predictive reference trajectory over the control horizon $T_{\mathrm{fut}}$, a predictive LP filter computes the next $T_{\mathrm{fut}}$ samples of the desired yaw angle $\psi_d$ from the commanded reference $\psi_{\mathrm{ref}}$.
At each control step, DeePC determines the optimal sequence of rudder deflections $\delta_r$ over the prediction horizon by minimizing the objective function defined in~\eqref{eq:DeePC}. Only the first control action is applied to the system, after which the resulting yaw angle $\psi$ is measured. The applied control input $\delta_r$ and the measured output $\psi$ are then appended to the vectors $u_{\mathrm{past},\delta_r}$ and $y_{\mathrm{past},\psi}$, which are required to estimate the implicit current system state through a linear combination of previously collected trajectories.
This offline dataset used for constructing the data matrices $U_d$ and $Y_d$ was generated in the REMUS~100 AUV simulator by exciting the rudder input $\delta_r$ with a chirp signal overlaid with a pseudo-random binary sequence (PRBS), while recording the corresponding yaw response $\psi$. Since both signals operate within the same dynamic mode, the resulting dataset is sufficiently rich and satisfies the persistency-of-excitation condition. To capture a larger number of data points and therefore extend the operational range represented in the data, the Page matrix, given in~\eqref{eq:mosaic-Hankelmatrix}, was adopted instead of the Hankel matrix. This receding-horizon loop is then executed iteratively throughout the control process.
\subsection{DeePC Depth Control}
\begin{figure}[ht]
\begin{center}
\includegraphics[width=\linewidth]{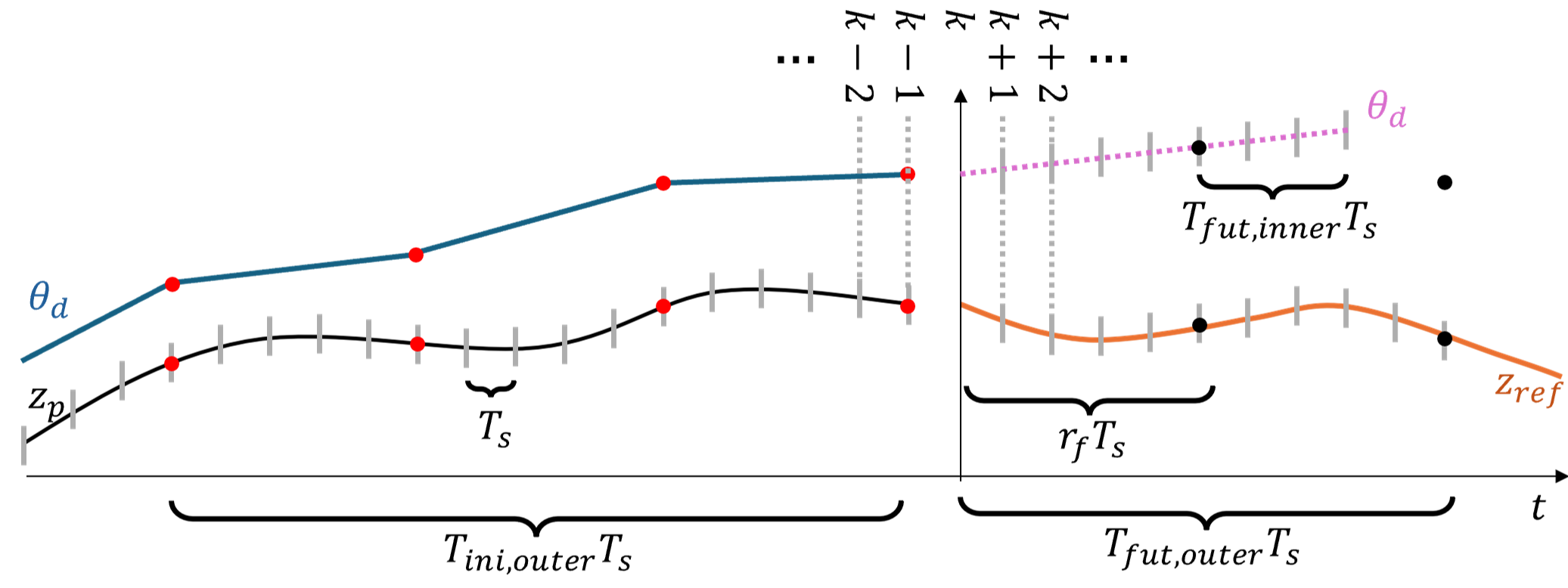}   
\caption{Illustration of the sampled points (red and black dots) along the past and future trajectories from the outer loop to illustrate the reference trajectory generation for the inner control loop in pink via linear extrapolation.} 
\label{fig:Sampling}
\end{center}
\end{figure}

To integrate DeePC into the depth control, two possible control architectures can be considered. The first approach directly maps the stern-plane deflection $\delta_s$ to the heave position $z_p$. However, since the depth response of an AUV exhibits significantly slower dynamics compared to the control input $\delta_s$, the system effectively behaves as a strong LP filter. This leads to two major issues: obtaining persistently exciting data becomes difficult due to the limited dynamic coupling between input and output, and reducing the control frequency to be able to collect sufficiently rich data may cause higher-frequency disturbances to remain unregulated.
Therefore, a cascaded DeePC structure, analogous to the classical depth control scheme, is adopted. In this configuration, the outer DeePC maps the desired heave position $z_{p,d}$ to a desired pitch angle $\theta_d$, while the inner DeePC regulates $\theta_d$ through the stern-plane deflection $\delta_s$. The inner loop design closely follows the heading control formulation discussed previously.
Due to the cascaded DeePC structure, the two control loops operate at different frequencies, with the inner loop running $r_f = f_{\mathrm{inner}} / f_{\mathrm{outer}}$ times faster than the outer loop. This allows the inner DeePC to compensate for disturbances with faster dynamics, while the outer loop tracks the heave position trajectory. Because the pitch dynamics lie between the fast dynamics of $\delta_s$ and the slow depth dynamics, this cascaded structure ensures both sufficient excitation and improved robustness. More generally, the loop-frequency separation extends the applicability of DeePC to plants whose dominant output channel evolves significantly slower than the actuator bandwidth, a regime where direct DeePC application is precluded by the persistency-of-excitation requirement within feasible data lengths.\\
While the data collection process for the inner loop is straightforward as for the heading control, collecting suitable data for the outer loop requires additional consideration, as its effective plant consists of the inner DeePC controller combined with the AUV dynamics. At first glance, this appears incompatible with the DeePC framework, as the inner DeePC involves an embedded optimization, breaking the deterministic LTI structure typically assumed by Willems’ Fundamental Lemma. To overcome this, the inner loop is replaced by a unit gain under the unity-gain assumption from successive loop closure design~\citep{Beard_2012}, meaning that the pitch angle is treated as tracking its reference exactly, i.e., $\theta = \theta_d$. This allows the pitch angle $\theta$ to be used as the effective input and the measured heave position $z_p$ as the output for the outer-loop data sequences. \\
Two strategies can be considered for collecting the outer-loop data. The first subsamples $\theta$ and $z_p$ every $r_f$-th step from the inner-loop dataset. This requires the inner-loop excitation $\delta_s$ to simultaneously cover the pitch dynamics range for the inner DeePC and the depth dynamics range for the outer DeePC, which is impractical given the disparate timescales of the two channels. The preferred two-stage procedure separates the excitation by stage: the inner-loop data are collected first by exciting $\delta_s$, then the inner DeePC is closed during the outer-loop data collection while a $\theta_d$-chirp overlaid with PRBS at the outer-loop frequency excites the depth channel through the closed inner loop. Each stage's excitation therefore targets its own dynamic range, ensuring persistency of excitation in both channels. The outer-loop sequences are stored in the Page matrix format to extend the operational range covered by the data.\\
During control execution, the inner DeePC loop operates at the higher inner-loop frequency $f_{\mathrm{inner}}$, which matches the AUV simulation and sampling frequency, while the outer loop runs at the lower frequency $f_{\mathrm{outer}}$ with decimation $r_f = f_{\mathrm{inner}}/f_{\mathrm{outer}}$. When the outer loop is executed at time step $k$, the predictive LP filter generates a heave-position reference trajectory of $T_{\mathrm{fut}}$ samples at the outer-loop rate, which serves as the reference for the outer DeePC optimization. The past-data vectors $u_{\mathrm{past},\theta}$ and $y_{\mathrm{past},z_p}$ contain the most recent $T_{\mathrm{ini}}$ outer-loop samples of $\theta$ and $z_p$ at the same outer-loop rate. The outer DeePC then computes the predicted desired-pitch sequence $\theta_{d,k+r_f}, \ldots, \theta_{d,k+T_{\mathrm{fut}} r_f}$, of which only the first element $\theta_{d,k+r_f}$ is applied as the inner-loop reference following the receding-horizon principle.
Between successive outer-loop updates, $\theta_d$ is linearly interpolated to ensure smooth transitions and reduce control jitter, and extrapolated up to $k + r_f + T_{\mathrm{fut,inner}}$ so all inner-loop DeePC optimizations have access to a continuous reference (Fig.~\ref{fig:Sampling}, pink dashed line).
\begin{table}[h]
\centering
\caption{Configuration of the three DeePC instances in the proposed cascaded framework.}
\label{tab:DeePC_instances}
\setlength{\tabcolsep}{4pt}
\resizebox{\columnwidth}{!}{%
\begin{tabular}{lcccc}
\hline
DeePC instance & Input & Output & Reference (source) & Rate \\
\hline
Heading & $\delta_r$ & $\psi$ & $\psi_d$ (LP filter) & $f_{\mathrm{inner}}$ \\
Outer depth & $\theta$ & $z_p$ & $z_{p,d}$ (LP filter) & $f_{\mathrm{outer}}$ \\
Inner depth & $\delta_s$ & $\theta$ & $\theta_d$ (outer DeePC) & $f_{\mathrm{inner}}$ \\
\hline
\end{tabular}%
}
\end{table}

\section{Predictive Adaptive Line-of-Sight}
\label{sec:palos}
\begin{figure}[ht]
\begin{center}
\includegraphics[width=\linewidth]{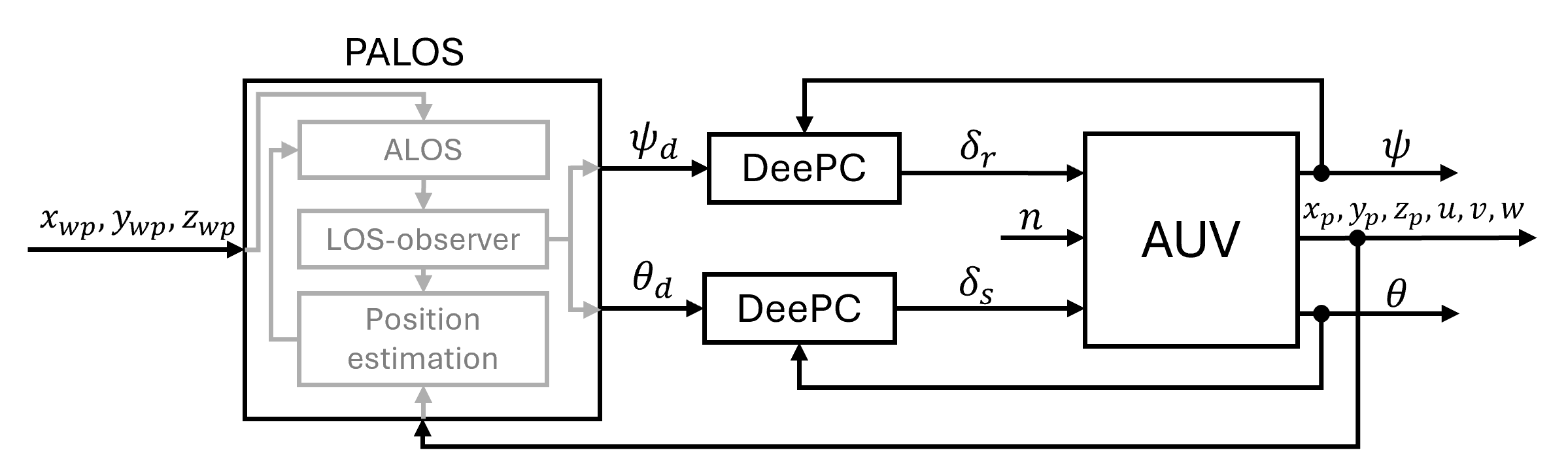}    
\caption{Integration of PALOS into the DeePC framework for 3-D Path following.} 
\label{fig:PALOS_DeePC}
\end{center}
\end{figure}

For AUV missions where the reference trajectory must be generated from a spatial waypoint sequence rather than prescribed directly, the proposed framework employs a predictive guidance layer that supplies a horizon-consistent reference trajectory to the DeePC tracking layer, enabling 3-D waypoint tracking with the receding-horizon predictive control structure. The guidance is built on the recent ALOS algorithm of Fossen and Aguiar~\citep{Fossen_2024}, which provides drift-compensated heading and pitch commands for AUV path following but produces only a single-step output incompatible with the horizon-length input required by receding-horizon predictive controllers such as DeePC. This section presents an extension of ALOS to a predictive multistep formulation, denoted PALOS guidance, that acts as a trajectory generator producing the full prediction-horizon reference consumed by the DeePC tracking layer. The integration of PALOS with DeePC is illustrated in Fig.~\ref{fig:PALOS_DeePC}.
The spatial path is described by the position vector $p_s(s) = [x_s(s), y_s(s), z_s(s)]^\top$, parameterized by the path variable $s$ and simplified in this work to a straight-line segment between two consecutive waypoints in the NED frame. The local tangent vector $t_t(s)$ between active waypoints $p_{wp,k} = [x_{wp,k}, y_{wp,k}, z_{wp,k}]^\top$ and $p_{wp,k+1} = [x_{wp,k+1}, y_{wp,k+1}, z_{wp,k+1}]^\top$ in the NED frame is
\begin{equation}
t_t(s) = \begin{bmatrix} x'_t \\ y'_t \\ z'_t \end{bmatrix} = \begin{bmatrix} x_{wp,k+1} - x_{wp,k} \\ y_{wp,k+1} - y_{wp,k} \\ z_{wp,k+1} - z_{wp,k} \end{bmatrix}.
\label{eq:tangent}
\end{equation}
The associated path-tangential frame $\{p\}$ has its origin at $p_{wp,k}$ and its $x^p$-axis pointing toward $p_{wp,k+1}$. The path heading and elevation angles follow as
\begin{equation}
\pi_h = \arctan\!\left(\frac{y'_t}{x'_t}\right), \; \pi_v = \arctan\!\left(\frac{-z'_t}{\sqrt{(x'_t)^2 + (y'_t)^2}}\right).
\label{eq:pathangles}
\end{equation}
The along-, cross-, and vertical-track errors $(x^p_e, y^p_e, z^p_e)$ in $\{p\}$ are obtained by rotating the NED position error $p^n_e = p - p_{wp,k}$, where $p = [x_p, y_p, z_p]^\top$ is the current AUV position vector and $p_{wp,k}$ is the active waypoint serving as the origin of $\{p\}$:
\begin{equation}
\scalebox{0.91}{$
\begin{bmatrix}x_e^p\\y_e^p\\z_e^p\end{bmatrix}
\!=\!
\begin{bmatrix}
\cos\pi_v & 0 & \sin\pi_v\\
0 & 1 & 0\\
-\sin\pi_v & 0 & \cos\pi_v
\end{bmatrix}^{\!\top}\!
\begin{bmatrix}
\cos\pi_h & -\sin\pi_h & 0\\
\sin\pi_h & \cos\pi_h & 0\\
0 & 0 & 1
\end{bmatrix}^{\!\top}\!
\begin{bmatrix}x_e^n\\y_e^n\\z_e^n\end{bmatrix}
\label{eq:errorframe}
$}
\end{equation}
The ALOS guidance law then commands the desired heading $\psi_d$ and pitch $\theta_d$ with adaptive compensation for the unknown drift-induced crab and angle-of-attack:
\begin{equation}
\begin{aligned}
&\psi_d = \pi_h - \hat{\beta}_c - \arctan\!\left(\frac{y_e^p}{\Delta_h}\right), \\
&\theta_d = \pi_v + \hat{\alpha}_c + \arctan\!\left(\frac{z_e^p}{\Delta_v}\right),
\label{eq:ALOS_angles}
\end{aligned}
\end{equation}
where $\Delta_h, \Delta_v > 0$ are the horizontal and vertical look-ahead distances. The adaptive estimates $\hat{\beta}_c$ and $\hat{\alpha}_c$ are updated through the projection-modified adaptation laws
\begin{equation}
\dot{\hat{\beta}}_c = \gamma_h \, \frac{\Delta_h \, y^p_e}{\sqrt{\Delta_h^2 + (y^p_e)^2}}, \qquad \dot{\hat{\alpha}}_c = \gamma_v \, \frac{\Delta_v \, z^p_e}{\sqrt{\Delta_v^2 + (z^p_e)^2}},
\label{eq:alos_adapt}
\end{equation}
with positive adaptation gains $\gamma_h, \gamma_v > 0$. In implementation the right-hand sides are passed through the parameter projection of~\citep{Fossen_2024} to enforce $|\hat{\beta}_c| \leq M_\theta$ and $|\hat{\alpha}_c| \leq M_\theta$ for a bound $M_\theta > 0$, preserving the stability properties under estimation transients. Under straight-line path following at constant cruise speed, the origins of the cross- and vertical-track error dynamics are uniform semiglobal exponentially stable, with theoretical guarantees established in~\citep{Fossen_2024}. The raw commanded angles $\psi_d, \theta_d$ are filtered through a discrete-time LOS observer that produces smooth, differentiable reference signals along with their rate commands $r_d, q_d$, suppressing discontinuities at waypoint transitions for the inner control loops.

PALOS extends ALOS to a multistep formulation in which the AUV state is recursively propagated forward across the prediction horizon, ALOS is re-evaluated at each predicted position, the LOS observer is applied to filter the reference, and the active waypoint pair is advanced whenever a predicted position reaches the next waypoint. The procedure produces a horizon-length sequence of desired heading and pitch angles, $\{\psi_d^{(i)}, \theta_d^{(i)}\}_{i=1}^{T_{\mathrm{fut}}}$, that serves as the trajectory-generation output for the DeePC tracking layer.
The forward propagation is built on the 3-D amplitude-phase form of the NED kinematic differential equations introduced in~\citep{Fossen_2024}, which expresses the NED position rates $\dot{x}_p, \dot{y}_p, \dot{z}_p$ in terms of directed horizontal and vertical speed magnitudes $U_h$ and $U_v$, respectively:
\begin{equation}
\begin{aligned}
&\dot{x}_p = U_h \cos(\psi + \beta_c), \quad \\
&\dot{y}_p = U_h \sin(\psi + \beta_c), \quad \\ 
&\dot{z}_p = -U_v \sin(\theta - \alpha_c),
\end{aligned}
\label{eq:fossen_phase}
\end{equation}
where the speed magnitudes are derived from the body-frame surge velocity and the crab and angle-of-attack as
\begin{equation}
\begin{aligned}    
&U_v = u \sqrt{1 + \tan^2(\alpha_c)}, \qquad \\
&U_h = U_v \cos(\theta - \alpha_c) \sqrt{1 + \tan^2(\beta_c)},
\end{aligned}
\label{eq:Uhv}
\end{equation}
and the actual crab $\beta_c$ and angle-of-attack $\alpha_c$ relate to the body-frame velocities and roll through
\begin{equation}
\begin{aligned}
&\alpha_c = \arctan\!\left(\frac{v \sin\phi + w \cos\phi}{u}\right), \qquad \\
&\beta_c = \arctan\!\left(\frac{v \cos\phi - w \sin\phi}{U_v \cos(\theta - \alpha_c)}\right).
\end{aligned}
\label{eq:crab_aoa}
\end{equation}
Under the assumption that the body-frame velocities $(u, v, w)$ and the roll $\phi$ remain approximately constant across the prediction horizon, consistent with the fixed propeller speed $n_p$ established in Section~\ref{sec:classic} and the steady-cruise regime of straight-line path following, the magnitudes $U_h$ and $U_v$ in~\eqref{eq:Uhv} become constants over the horizon and are computed once at the start of each PALOS evaluation. Assuming exact tracking over the prediction horizon, as standard in receding-horizon approaches, $\psi = \psi_d$, and $\theta = \theta_d$ follow. The discrete-time PALOS propagation from prediction step $i$ to step $i{+}1$ takes then the form
\begin{equation}
\begin{aligned}
x_p^{(i+1)} &= x_p^{(i)} + h \, U_h \cos\!\big(\psi_d^{(i)} + \hat{\beta}_c^{(i)}\big), \\
y_p^{(i+1)} &= y_p^{(i)} + h \, U_h \sin\!\big(\psi_d^{(i)} + \hat{\beta}_c^{(i)}\big), \\
z_p^{(i+1)} &= z_p^{(i)} - h \, U_v \sin\!\big(\theta_d^{(i)} - \hat{\alpha}_c^{(i)}\big),
\end{aligned}
\label{eq:palos_propagation}
\end{equation}
where $h$ is the sampling period and the adaptive estimates $\hat{\beta}_c^{(i)}, \hat{\alpha}_c^{(i)}$ substitute for the true crab and angle-of-attack in the absence of direct measurement. These estimates evolve along the prediction horizon through the adaptation law in~\eqref{eq:alos_adapt}, which is re-applied at every recursive ALOS evaluation step. The propagation is therefore horizon-coherent with the adaptive drift compensation predicted to evolve along the trajectory: the predicted course over ground $\psi_d^{(i)} + \hat{\beta}_c^{(i)}$ and flight-path angle $\theta_d^{(i)} - \hat{\alpha}_c^{(i)}$ track the predicted physical motion direction in receding horizon manner. Only the one-step-ahead estimates $\hat{\beta}_c^{(1)}, \hat{\alpha}_c^{(1)}$ are committed to the persistent real-time state used as the initial condition of the next prediction horizon.

When a predicted position $p^{(i+1)} = [x_p^{(i+1)}, y_p^{(i+1)}, z_p^{(i+1)}]^\top$ within the prediction horizon enters a sphere of acceptance with radius $R_{\mathrm{switch}}$ around the next waypoint, the waypoint index is advanced from that prediction step onward, and subsequent prediction steps within the same horizon evaluate ALOS with respect to the new updated waypoint. As with the crab-angle estimates, only the one-step-ahead waypoint index is committed to the real-time state.

The resulting reference trajectory is horizon-consistent in the sense that each sample of $\psi_d^{(i)}$ and $\theta_d^{(i)}$ over the prediction horizon is generated from the ALOS guidance geometry evaluated at the predicted AUV position at that step, rather than from the current AUV position alone. This property is what allows direct integration of the guidance layer into the DeePC optimization for 3-D waypoint tracking, without horizon-mismatch artifacts between the guidance output and the predictive-controller input. The PALOS multistep propagation is constructed directly on Fossen's amplitude-phase formulation, placing the predictive guidance layer on the same kinematic framework on which the uniform semiglobal exponential stability of the underlying ALOS cross- and vertical-track error dynamics is established~\citep{Fossen_2024}, providing a principled basis for the predictive guidance layer.

\section{Results}
\label{sec:results}
This section presents the simulation results obtained for the proposed DeePC-based control framework and compares its performance with the classical PI/PID control architecture across various scenarios, including nominal operation, external disturbances, operation beyond the data regime and 3-D path following. All scenarios are evaluated on the full nonlinear 6-DOF AUV simulation model described in Section~\ref{sec:auv}, with velocity-dependent damping, Coriolis-centripetal coupling, and attitude-dependent restoring forces active throughout.
For the classical architecture in all following scenarios, the best-performing PI/PID gains were selected from the preinitialized MSS toolbox for depth control and the tuning proposed in~\citep{Fossen_2023} for heading control. 
The PI/PID gains are as follows: 
\begin{equation*}
    \begin{aligned}
&K_{p,\theta} = 5.0,                               
&K_{i,\theta} = 0.3,                
&&K_{d,\theta} = 2.0, 
&&&K_{p,\psi} = 7.5, \\
&K_{i,\psi} = 0.75, 
&K_{d,\psi} = 15, 
&&K_{p,z} = 0.1, 
&&&K_{i,z} = 10^{-3}
    \end{aligned}
\end{equation*}
For DeePC, the inner and outer control loops operated at frequencies of $20\,\mathrm{Hz}$ and $2\,\mathrm{Hz}$, respectively, while for the PI/PID-based control, all loops operated at $20\,\mathrm{Hz}$. 
The maximum possible angle for both, stern and rudder plane, are $\delta_{s,max}=\delta_{r,max}=\pm20^\circ$, while the maximum allowed pitch angle of the AUV is $\theta_{max}=\pm30^\circ$. These limitations were satisfied in the dynamic model via a saturation function and included in the DeePC framework as the regarding input and output constraints, as in~\eqref{eq:DeePC}. 

\subsection{General Control Performance}
\label{sec:results_general}
\begin{figure}[ht]
    \centering
    \includegraphics[width=\linewidth]{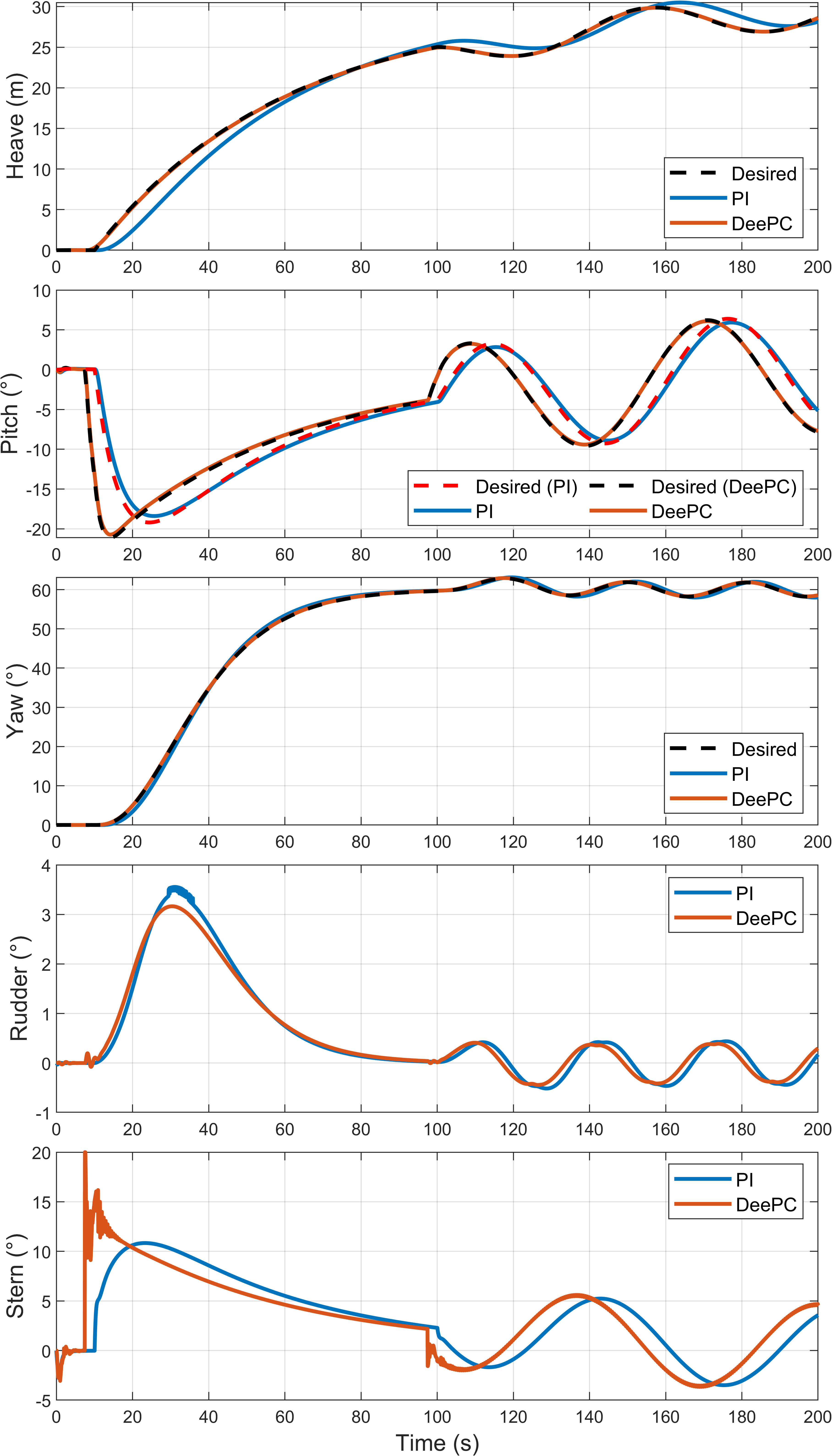}
    \caption{Responses in heave position $z_p$, pitch angle $\theta$, and yaw angle $\psi$ of DeePC and PI/PID to each desired reference. Associated control command angles for the rudder plane and stern plane, $\delta_r$ and $\delta_s$.}
    \label{fig:comparison_results}
\end{figure}

The PI/PID and DeePC-based controllers for heading and depth control are compared, with reference commands smoothed using the built-in LP filters of the MSS toolbox~\citep{fossen2004mss}. At $t = 10\,\mathrm{s}$, a step command was applied toward a depth of $30\,\mathrm{m}$ and a heading of $60^{\circ}$. At $t = 100\,\mathrm{s}$, a sinusoidal signal was superimposed on both reference trajectories to evaluate performance under more dynamic, albeit less realistic, reference variations.  
In both control approaches, the propeller speed was fixed to $n = 1000\,\mathrm{rpm}$ during data collection and control execution.  
The specific DeePC hyperparameters (HP) are listed in Table~\ref{tab:HPs}.\\
Figure~\ref{fig:comparison_results} illustrates a clear performance advantage of the DeePC controller, particularly in the heave control, due to its predictive nature in contrast to the reactive behavior of the PI/PID controller. This difference becomes especially evident when the sinusoidal reference is active, as the classical controller exhibits increased phase lag and tracking error. Quantitatively, the root-mean-square error (RMSE) values in Table~\ref{tab:RMSEs} confirm the superior accuracy of the DeePC approach.\\
\textit{Remark:} The different $\theta_d$ trajectories shown in Fig.~\ref{fig:comparison_results} for DeePC and PI/PID arise because each outer-loop controller computes its own pitch reference from the depth error.\\
Analysis of the control inputs, i.e., the rudder and stern-plane commands $\delta_r$ and $\delta_s$, respectively, reveals that the DeePC outputs exhibit abrupt changes at reference transitions. This behavior arises from the absence of modeled actuator dynamics in the employed AUV model. Incorporating these dynamics into the system model is therefore identified as a priority for future work to ensure physically realizable control signals. When data are collected from a real AUV, such actuator dynamics will naturally be embedded in the recorded sequences. The classical PI/PID controller, by contrast, produces smoother input transitions due to its inherent feedback structure and the LP filtering of the reference signals.
\setlength{\tabcolsep}{3pt} 
\begin{table}[!t]
\caption{RMSEs for the yaw angle $\psi$, heave position $z_p$, and pitch angle $\theta$ regarding the given references for the different scenarios. }
\label{tab:RMSEs}
\resizebox{\columnwidth}{!}{%
\begin{tabular}{l|cc|cc|cc|cc}
                        & \multicolumn{2}{c|}{8.1}                      & \multicolumn{2}{c|}{8.2}                  & \multicolumn{2}{c|}{8.3}                 & \multicolumn{2}{c}{8.4}                      \\
                        & \multicolumn{1}{c|}{DeePC} & \multicolumn{1}{c|}{PID} & \multicolumn{1}{c|}{DeePC} & \multicolumn{1}{c|}{PID} & \multicolumn{1}{c|}{DeePC} & \multicolumn{1}{c|}{PID} & \multicolumn{1}{c|}{DeePC} & \multicolumn{1}{c}{PID} \\ \hline
${\psi}_{RMSE} \,(^\circ)$   & \multicolumn{1}{c|}{0.04}      &    0.66                      & \multicolumn{1}{c|}{0.05}      &           0.66               & \multicolumn{1}{c|}{0.07}      &      0.7                    & \multicolumn{1}{c|}{0.12}  & 1.12                    \\
${z_{p}}_{RMSE}\,(m)$           & \multicolumn{1}{c|}{0.07}      &     1.27                     & \multicolumn{1}{c|}{0.06}      &             1.14             & \multicolumn{1}{c|}{3.17}      &        6.50                  & \multicolumn{1}{c|}{-}     & \multicolumn{1}{c}{-}   \\
${\theta}_{RMSE} \,(^\circ)$ & \multicolumn{1}{c|}{0.18}      &    0.62                      & \multicolumn{1}{c|}{0.13}      &            0.54             & \multicolumn{1}{c|}{9.60}      &            33.71              & \multicolumn{1}{c|}{0.012} & 0.21      
\end{tabular}
}
\end{table}
\subsection{Robustness Evaluation under Ocean Currents}
\label{sec:Currents}
To assess the robustness of the proposed DeePC-based control architecture, a stress test was conducted by introducing ocean current disturbances modeled as described in Appendix~\ref{app:ocean_current_model}.   
The horizontal current had a mean velocity magnitude of $V_c = 0.5\,\mathrm{m/s}$ acting in the direction $\beta_{V_c} = 150^\circ$ in the NED frame, while the vertical component was directed downward with $W_c = 0.05\,\mathrm{m/s}$. To emulate small-scale turbulence, zero-mean Gaussian perturbations were added at each time step to both the velocity and direction components of the current.  
Despite these ocean currents, which couple nonlinearly into the surge and sway damping via the relative velocity in~\eqref{eq:Damping}, the DeePC controller exhibited consistently robust behavior, maintaining accurate trajectory tracking in both the heading and depth channels. Quantitatively, the RMSE values presented in Table~\ref{tab:RMSEs} confirm the superior control performance under disturbances of the DeePC-based control compared to the conventional PI/PID architecture. \\
\textit{Remark: }Due to the downward-directed current, the vertical disturbance supports reaching the smoothed step reference slightly faster, which results in a lower ${z_p}_{RMSE}$ and $\theta_{RMSE}$ in Table~\ref{tab:RMSEs}, compared to Section~\ref{sec:results_general}.

\subsection{Robustness Beyond the Data Regime}
To assess the robustness of the DeePC-based control architecture under operating conditions beyond the data collection range, the AUV’s propeller speed was systematically reduced, thereby decreasing the surge, sway, and heave velocities. Since the DeePC data sequences were collected at $n = 1000\,\mathrm{rpm}$, this experiment effectively tests the controller’s generalization capability beyond its data regime. The vehicle is nonlinear across the entire operating envelope (Section~\ref{sec:auv}), and the velocity-dependent damping in surge and sway in~\eqref{eq:Damping} introduces additional nonlinear effects whose magnitude varies with relative velocity, so this scenario probes a regime where these effects diverge from the data collection point.
Generally, at reduced velocities, the AUV is physically unable to follow the reference profiles generated by the LP filter. Consequently, the classical PI/PID controller exhibits significant integral wind-up, resulting in large overshoot, rendering the RMSE values inapplicable for fair comparison. Although anti-windup schemes could mitigate this issue, they were not implemented here, since the DeePC formulation handles input and output constraints natively as in~\eqref{eq:DeePC}, and adding anti-windup to the PI/PID side would introduce asymmetric engineering effort rather than a fairer comparison.
In contrast, the DeePC controller demonstrated strong robustness to these nonlinear effects. Despite transient tracking degradation when the vehicle momentarily lacked sufficient speed due to the decreased propeller speed, the controller successfully recovered and maintained trajectory tracking once the reference flattened and the AUV could catch up to the reference. The RMSE value reported in Table~\ref{tab:RMSEs} corresponds to a propeller speed reduction to $700\,\mathrm{rpm}$.  
Further testing showed that DeePC achieved stable and accurate control up to the maximum propeller speed of $1525\,\mathrm{rpm}$ and maintained satisfactory performance down to approximately $300\,\mathrm{rpm}$ when the reference step was appropriately scaled. Below $300\,\mathrm{rpm}$, however, the nonlinearities exceeded the DeePC's depth-tracking capability. This operating point thus marks the observed nonlinearity-handling limit of the DeePC controller, which will be addressed in future work through adaptive or gain-scheduled extensions.

\subsection{3-D Path following}
\label{sec:3D-simu}
\begin{figure}[ht]
    \centering
    \includegraphics[width=\linewidth]{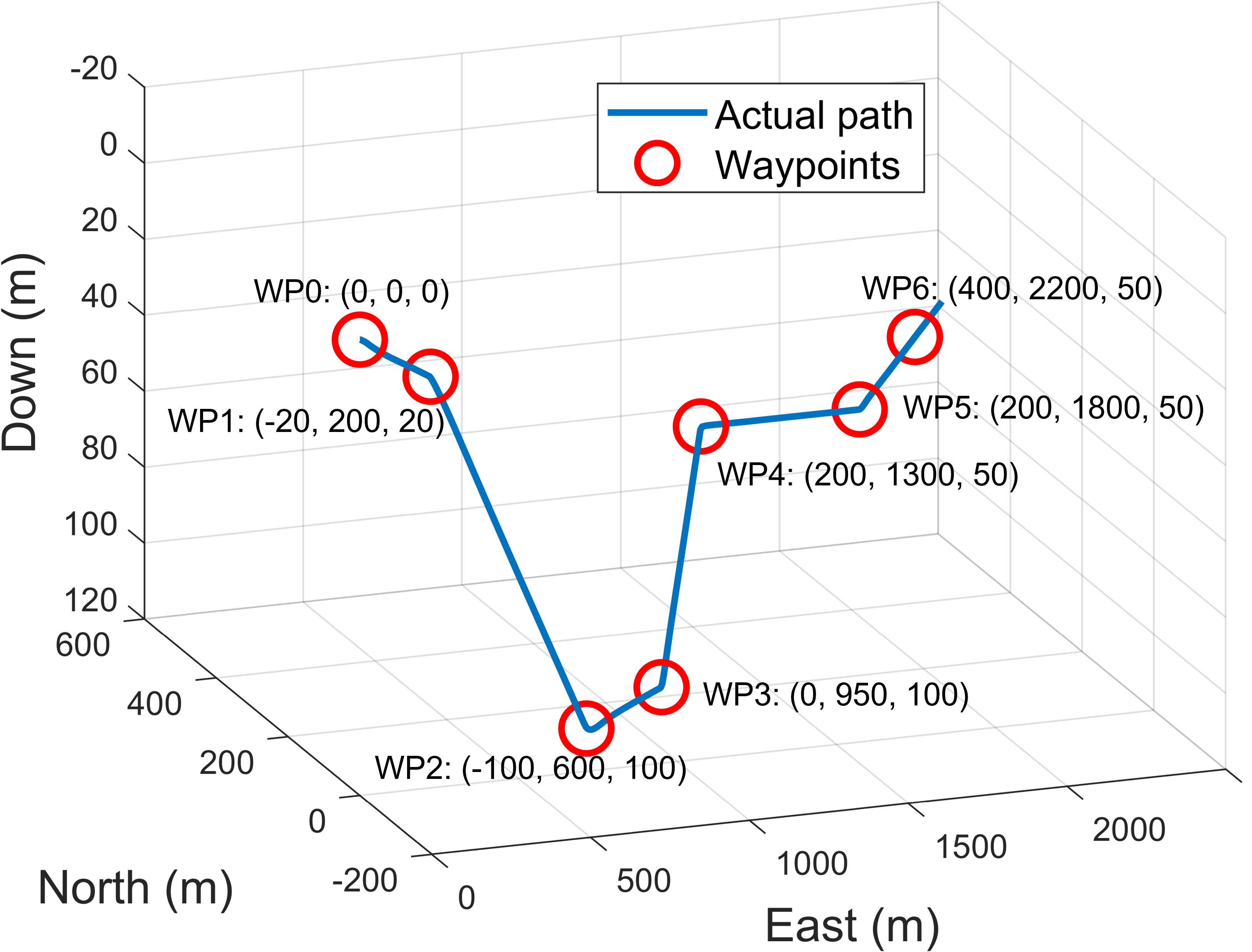}
    \caption{3-D path of the controlled AUV with desired waypoints in NED.}
    \label{fig:3D_Path}
\end{figure}

The proposed PALOS combined with DeePC was evaluated for 3-D waypoint-based path following. The propeller speed was fixed at $n = 1000\,\mathrm{rpm}$, and the waypoint coordinates $(x_{wp}, y_{wp}, z_{wp})$ are listed in Fig.~\ref{fig:3D_Path}. Disturbances are disabled in this scenario to isolate the guidance accuracy. The $\theta$-DeePC controller employed more aggressive hyperparameters, as the inner loop receives a continuously updated reference in every control iteration, effectively eliminating stern-plane jitter and enabling more aggressive $\theta$-DeePC. The corresponding HPs are summarized in Table~\ref{tab:HPs}.
The PALOS-DeePC architecture achieves accurate path tracking, successfully reaching all waypoints as shown in Fig.~\ref{fig:3D_Path}. At the scale of the experimental waypoint sequence, the ALOS-PI/PID trajectory is visually indistinguishable from the PALOS-DeePC trajectory, so the figure is included only as a visual reference for the 3-D waypoint-tracking task rather than as a comparative plot. The quantitative comparison appears in Table~\ref{tab:RMSEs}. The combined PALOS-DeePC system reduces the inner-loop tracking errors by roughly an order of magnitude relative to ALOS-PI/PID, from $\psi_{RMSE} = 1.12^\circ$ and $\theta_{RMSE} = 0.21^\circ$ to $0.12^\circ$ and $0.012^\circ$ respectively. These reductions measure how accurately the inner controller tracks the guidance command and are therefore attributable to the data-driven inner-control layer, consistent with the inner-loop comparisons in Sections~\ref{sec:results_general} and~\ref{sec:Currents}. The horizontal cross-track error reduces in root-mean-square (RMS) magnitude from $0.28\,\mathrm{m}$ for ALOS-PI/PID to $0.20\,\mathrm{m}$ for PALOS-DeePC, a reduction of approximately $28\%$, and the vertical-track RMS reduces from $0.10\,\mathrm{m}$ to $0.09\,\mathrm{m}$. 
This improvement reflects the same inner-loop attribution as the heading and pitch reductions. Both guidance laws share the same ALOS geometry, and PALOS extends ALOS only as a multistep prediction along the horizon. The two AUVs occupy different positions over time as a consequence of the inner controllers tracking the commands more or less accurately. The cross-track difference between the two combined systems is therefore set by the inner-loop choice, not by a difference in the guidance laws themselves. The intended role of PALOS is to enable ALOS as a multistep predictive variant that composes with predictive controllers such as DeePC by supplying a horizon-consistent reference trajectory, not to outperform ALOS at the guidance layer.
At waypoint transitions, where the path direction changes abruptly, the ALOS-PI/PID combination exhibits higher cross-track errors that decay once the new path segment is reached, while the predictive PALOS-DeePC framework anticipates the transition and adjusts the control action preemptively, maintaining smoother tracking through the transition. 

\subsection{Hyperparameter Tuning Procedure}
\label{sec:HP}
The DeePC HPs were tuned in two stages: an initial coarse search varying each HP individually, followed by refinement with ten independent simulations per candidate value (five nominal, five with ocean-current disturbances) to account for PRBS stochasticity. The best-performing value at each stage was selected by tracking RMSE and qualitative control-input smoothness. The performance valley was relatively flat across all HPs, so the final values prioritize smoothness over marginal RMSE improvements. The selected HPs are summarized in Table~\ref{tab:HPs}; for the 3-D path-following case in Section~\ref{sec:3D-simu}, the $\theta$-DeePC controller uses $Q = 100$ and $R = 100$.
\setlength{\tabcolsep}{6pt} 
\begin{table}[ht]
\centering
\caption{Hyperparameters used for DeePC in the experiments. }
\label{tab:HPs}
\begin{tabular}{l|c|c|c|c|c|c|c}
             & $T_{fut}$ & $T_{ini}$ & $T_{d}$ & $\lambda_{y}$ & $\lambda_{g}$ & $Q$ & $R$ \\ \hline
$\psi$       & $10$      & $6$       & $100$   & $10^7$          & $10^3$        & $10^4$ & $0.1$ \\
$z_p$          & $7$       & $7$       & $200$   & $10^5$          & $10^2$        & $10^2$ & $10^3$ \\
$\theta$     & $6$       & $5$       & $100$   & $10^7$          & $10^2$        & $10^4$ & $500$ \\

\end{tabular}
\end{table}

\section{Conclusion and Further Work}
\label{sec:conc}
This work successfully demonstrated DeePC for AUV control in both heading and depth regulation in a high fidelity simulation environment. A cascaded DeePC architecture was developed for depth control, incorporating a loop-frequency separation concept that proved effective under disturbance and robustness tests. Furthermore, the PALOS algorithm was proposed and integrated with DeePC, enabling successful 3-D waypoint path following of the REMUS~100 AUV.
The results confirm that the proposed PALOS-DeePC approach constitutes a robust, data-driven, and predictive control solution, achieving high performance without requiring an explicit dynamic model and the associated effort. DeePC consistently outperformed the classical PI/PID architecture across all evaluated scenarios. The performance gap reflects the structural difference between predictive and reactive control, while the data-driven framework additionally eliminates the hydrodynamic identification effort that constrains parametric predictive alternatives such as MPC.\\
Future work will extend the approach towards more nonlinear operation across the full speed range with gain-scheduled DeePC formulations recently developed in \citep{Zieglmeier_2025}, and advancing from waypoint-based path following to full trajectory tracking. Furthermore, experimental validation on a real AUV is planned to demonstrate the method’s practical applicability.
\section*{DECLARATIONS}
\noindent\textbf{Conflict of Interest}\\
The authors declare that there is no competing financial interest or personal relationship that could have appeared to influence the work reported in this paper.\\
\textbf{Data Availability}\\
The source code used for the data collection, the DeePC implementations, and the experiments is available at \url{https://github.com/SebsDevLab/DeePC_AUV.git}.\\
\textbf{Declaration of Generative AI and AI Technologies in the Writing Process:}\\
During the preparation of this work, the author(s) used ChatGPT and Claude in order to improve the grammar of this paper. After using this tool/service, the author(s) reviewed and edited the content as needed and take full responsibility for the content of the publication.
\bibliographystyle{IEEEtran}
\bibliography{MyBib2}
\clearpage

\appendix
\section{Additional Notation}
\label{app:notation}
To simplify the representation of the AUV dynamics, the state vector is expressed with respect to two coordinate systems: the NED frame and the body-fixed frame. Using the Euler angle transformation, the rotation matrix \({R}_b^n\) defined in~\eqref{eq_app:Transformation-Matrix} transforms velocities from the body-fixed frame \(b\) to the NED frame \(n\), as shown in~\eqref{eq:Transformation}~\citep{fossen2021handbook}, where $v_n$ and $v_b$ are velocities with respect to the NED and body-fixed frames respectively. The inverse matrix \({R}_n^b = ({R}_b^n)^{-1}\) enables the transformation in the opposite direction. Here, \(c\psi\) and \(s\psi\) denote the cosine and sine of the angle \(\psi\), respectively.
\begin{equation}
{v}_n={R}_b^n {v}_b 
\label{eq:Transformation}
\end{equation}
\begin{equation}
\resizebox{0.9\columnwidth}{!}{%
${R}_b^{n}=\left[\begin{array}{ccc}
\mathrm{c} \psi \mathrm{c} \theta & -\mathrm{s} \psi \mathrm{c} \phi+\mathrm{c} \psi \mathrm{~s} \theta \mathrm{~s} \phi & \mathrm{~s} \psi \mathrm{~s} \phi+\mathrm{c} \psi \mathrm{c} \phi \mathrm{~s} \theta \\
\mathrm{~s} \psi \mathrm{c} \theta & \mathrm{c} \psi \mathrm{c} \phi+\mathrm{s} \phi \mathrm{~s} \theta \mathrm{~s} \psi & -\mathrm{c} \psi \mathrm{~s} \phi+\mathrm{s} \theta \mathrm{~s} \psi \mathrm{c} \phi \\
-\mathrm{s} \theta & \mathrm{c} \theta \mathrm{~s} \phi & \mathrm{c} \theta \mathrm{c} \phi
\end{array}\right]$
}
\label{eq_app:Transformation-Matrix}
\end{equation}
The angular velocity mapping matrix $T_k$ maps body angular velocities to Euler-angle rates. 
\begin{equation}
\resizebox{0.5\columnwidth}{!}{%
$T_k =
\begin{bmatrix}
1 & \sin\phi\tan\theta & \cos\phi\tan\theta \\[4pt]
0 & \cos\phi           & -\sin\phi           \\[4pt]
0 & \sin\phi/\cos\theta & \cos\phi/\cos\theta
\end{bmatrix}$}
\label{eq_app:Matrix2}
\end{equation}
The skew-symmetric matrix $S(r) \in \mathbb{R}^{3\times3}$ associated with a vector $r = [r_x, r_y, r_z]^\top $ is defined as
\begin{equation}
S(r) =
\begin{bmatrix}
0 & -r_z & r_y \\
r_z & 0 & -r_x \\
-r_y & r_x & 0
\end{bmatrix}.
\label{eq_app:Skew-matrix}
\end{equation}
The system transformation matrix $H(r)$, defined for a vector $r \in \mathbb{R}^3$, is given by~\eqref{eq_app:H_matrix}, where $I_{3\times3}$ is the identity matrix, $0_{3\times3}$ is the zero matrix.
\begin{equation}
H(r) =
\begin{bmatrix}
I_{3\times3} & S(r)^\top  \\
0_{3\times3} & I_{3\times3}
\end{bmatrix}
\label{eq_app:H_matrix}
\end{equation}

\section{Parameter Introduction for AUV dynamic matrices}
\label{app:intro}
Prior to describing the components of the 6-DOF rigid-body equations of motion, some fundamental concepts regarding the vehicle geometry must be established. The REMUS~100 AUV is represented using the Myring hull equations~\citep{Myring_1976} and, in the MSS toolbox, approximated as a spheroid~\citep{fossen2004mss}. The spheroid is defined by its longitudinal, transverse, and vertical semi-axes $a$, $b$, and $c$. Owing to the near-cylindrical shape of the vehicle, the transverse and vertical semi-axes are equal ($b = c$), and the geometry satisfies the relation in~\eqref{eq_app:spheroid} in terms of the AUV's length $L_{\mathrm{AUV}}$ and diameter $D_{\mathrm{AUV}}$. To ensure consistency with the vehicle mass, a rescaling of the semi-axes is performed according to~\eqref{eq_app:spheroid_mass}, where $m$ is the vehicle mass and $\rho$ the fluid density. Solving these relations yields the semi-axis values $a$ and $b$ defining the equivalent spheroidal representation of the REMUS~100.
\begin{equation}
    \frac{L_{\mathrm{AUV}}^2}{a^2} + 2\frac{D_{\mathrm{AUV}}^2}{b^2} = 1
    \label{eq_app:spheroid}
\end{equation}
\begin{equation}
    m = \frac{4}{3}\pi\rho a b^2
    \label{eq_app:spheroid_mass}
\end{equation}
Using the semi-axes $a$ and $b$ of the spheroidal approximation, the moments of inertia about the body-fixed axes $x_b$, $y_b$, and $z_b$ are computed according to~\eqref{eq_app:Ix}-\eqref{eq_app:Iz}.
\begin{equation}
    I_{x_b} = \frac{2}{5} m \, b^2
    \label{eq_app:Ix}
\end{equation}
\begin{equation}
    I_{y_b} = I_{z_b} = \frac{1}{5} m \, (a^2 + b^2)
    \label{eq_app:Iz}
\end{equation}
The vectors $r_{bG}, r_{bB} \in \mathbb{R}^3$ denote the positions of the vehicle’s center of gravity and buoyancy relative to the body-fixed origin. Using the transformation matrix $H(r)$ in~\eqref{eq_app:H_matrix}, forces and moments can be mapped between frames shifted by these vectors.

To estimate the added system inertia matrix the factors $\alpha_0$ and $\beta_0$ with eccentricity $e$ needs to be introduced: 
\begin{equation}
    e=1-\left(\frac{b}{a}\right)^2
\end{equation}
\begin{equation}
\alpha_0=\frac{2\left(1-e^2\right)}{e^3}\left(\frac{1}{2} \ln \frac{1+e}{1-e}-e\right)
\end{equation}
\begin{equation}
\beta_0=\frac{1}{e^2}-\frac{1-e^2}{2 e^3} \ln \frac{1+e}{1-e}
\end{equation}
Following the $k$-factors introduced in~\citep{Lamb_1924}, the added-mass contributions to the system inertia can be estimated. The factors are defined as
\begin{equation}
\begin{aligned}
&k_1 = \frac{\alpha_0}{2-\alpha_0}, \quad k_2 = \frac{\beta_0}{2-\beta_0}, \quad \\
&k^{\prime} = \frac{e^4(\beta_0-\alpha_0)}{(2-e^2)\left(2 e^2 - (2-e^2)(\beta_0-\alpha_0)\right)},
\end{aligned}
\end{equation}
yielding the diagonal added-mass matrix
\begin{equation}
\begin{aligned}
{M}_A &= -\operatorname{diag}\{X_{\dot{u}}, Y_{\dot{v}}, Z_{\dot{w}}, K_{\dot{p}}, M_{\dot{q}}, N_{\dot{r}}\} \\
&= \operatorname{diag}\{ m k_1, m k_2, m k_2, r_{44} I_x, k^{\prime} I_y, k^{\prime} I_y \}.
\end{aligned}
\label{eq_app:Added_M_M}
\end{equation}
While classical hydrodynamics predicts zero added moment in roll (element $(4,4)$ of $M_A$) for a spheroidal hull~\citep{Imlay_1961}, practical vehicles such as the REMUS~100 experience additional hydrodynamic contributions from the control surfaces and the propeller. Accordingly, this roll-related entry of the added-mass matrix is nonzero, $M_A(4,4)=r_{44} I_x$, capturing these effects in the simulation~\citep{fossen2004mss}.

The relative speed $U_r$ given in~\eqref{eq_app:Total_rel_speed} is defined as the Euclidean norm of the body-fixed linear velocity components of the AUV relative to the surrounding water, while the relative horizontal and vertical speeds, $U_{rh}$ and $U_{rv}$ respectively, are defined as in \eqref{eq_app:V_H_speed}.
\begin{equation}
U_r  = \sqrt{u_r^2 + v_r^2 + w_r^2}
\label{eq_app:Total_rel_speed}
\end{equation}
\begin{equation}
U_{rh} = \sqrt{u_r^2 + v_r^2}, \quad 
U_{rv} = \sqrt{u_r^2 + w_r^2}
\label{eq_app:V_H_speed}
\end{equation}
The vehicle's orientation relative to the oncoming flow can be characterized by the angle of attack $\alpha$ and the sideslip angle $\beta$, defined in terms of the relative velocity components as
\begin{equation}
\alpha = \tan^{-1}\!\left(\frac{w_r}{u_r}\right), 
\qquad
\beta = \sin^{-1}\!\left(\frac{v_r}{U_r}\right).
\label{eq_app:angle_of_attack}
\end{equation}

\section{The AUV dynamic matrices}
\label{app:dynamics}
The kinematic component of the 6-DOF rigid-body equations of motion for the AUV, given in~\eqref{eq:kinematic}, is defined by the transformation matrix $J(\eta)$. This matrix maps body-fixed linear and angular velocities to the time derivatives of the vehicle’s position and orientation in the NED frame, as detailed in~\eqref{eq_app:kin_Transformation}. It is constructed from the rotation matrix in~\eqref{eq_app:Transformation-Matrix} and the angular velocity transformation matrix in~\eqref{eq_app:Matrix2}.
\begin{equation}
J({\eta}) =
\begin{bmatrix}
{R_b^n}(\phi, \theta, \psi) & {0}_{3\times3} \\[6pt]
{0}_{3\times3} & {T_k}(\phi, \theta, \psi)
\end{bmatrix}
\label{eq_app:kin_Transformation}
\end{equation}
The inertia matrix $M$ of the AUV is composed of the rigid-body inertia $M_{RB}$ and the added-mass matrix $M_A$. The term $M_{RB}$ depends on the vehicle’s mass properties and is expressed through the transformation matrix $H$ from~\eqref{eq_app:H_matrix}, defined with respect to the center of gravity vector $r_{bG}$. The hydrodynamic contribution $M_A$, introduced in~\eqref{eq_app:Added_M_M}, accounts for the added inertia caused by the acceleration of the surrounding fluid. 
\begin{equation}
{M}_{\mathrm{RB}}={H}^{\top}\left({r}_{\mathrm{bg}}\right) \operatorname{diag}\left\{m, m, m, I_{x_b}, I_{y_b}, I_{z_b}\right\} {H}\left(r_{\mathrm{bg}}\right)
\label{eq_app:M_RB}
\end{equation}
In the MSS toolbox, $M_A$ is assumed constant, resulting in a constant $M$. The REMUS 100 exhibits both port–starboard and fore–aft symmetry, which simplifies the inertia representation according to~\citep{fossen2021handbook} to the sparse structured form shown in~\eqref{eq_app:Inertia}.
\begin{equation}
\scalebox{0.9}{$
{M}=M_{RB}+M_A=
\begin{bmatrix}
m_{11} & 0 & 0 & 0 & m_{15} & 0 \\
0 & m_{22} & 0 & m_{24} & 0 & 0 \\
0 & 0 & m_{33} & 0 & 0 & 0 \\
0 & m_{42} & 0 & m_{44} & 0 & 0 \\
m_{51} & 0 & 0 & 0 & m_{55} & 0 \\
0 & 0 & 0 & 0 & 0 & m_{66}
\end{bmatrix}
$}
\label{eq_app:Inertia}
\end{equation}
The Coriolis--centripetal matrix $C(\nu_r)$ represents gyroscopic and cross-coupling effects arising from linear and angular motion. It is composed of the rigid-body term $C_{RB}(\nu_r)$ and the added-mass term $C_A(\nu_r)$ as
\begin{equation}
C(\nu_r) = C_{RB}(\nu_r) + C_A(\nu_r).
\end{equation}
The rigid-body part $C_{RB}(\nu_r)$ is derived from $M_{RB}$ and depends on the angular velocities, while $C_A(\nu_r)$ depends on the relative linear velocities. 
\begin{equation}
\scalebox{0.87}{$
{C}_{\mathrm{RB}}\left({v}_{\mathrm{r}}\right)={H}^{\top}\left({r}_{\mathrm{bg}}\right)
\scalebox{0.8}{$
\left[\begin{array}{cccccc}
0 & -m r & m q & 0 & 0 & 0 \\
m r & 0 & -m p & 0 & 0 & 0 \\
-m q & m p & 0 & 0 & 0 & 0 \\
0 & 0 & 0 & 0 & I_z r & -I_y q \\
0 & 0 & 0 & -I_z r & 0 & I_x p \\
0 & 0 & 0 & I_y q & -I_x p & 0
\end{array}\right]
$}
{H}\left(r_{\mathrm{bg}}\right)
$}
\label{eq_app:Coriolis}
\end{equation}
\begin{equation}
\scalebox{0.8}{$
C_{\mathrm{A}}\left(v_r\right)=\left[\begin{array}{cccccc}
0 & 0 & 0 & 0 & -Z_{\dot{w}} w_r & Y_{\dot{v}} v_r \\
0 & 0 & 0 & Z_{\dot{w}} w_r & 0 & -X_{\dot{u}} u_r \\
0 & 0 & 0 & -Y_{\dot{v}} v_r & X_{\dot{u}} u_r & 0 \\
0 & -Z_{\dot{w}} w_r & Y_{\dot{v}} v_r & 0 & -N_{\dot{r}} r & M_{\dot{q}} q \\
Z_{\dot{w}} w_r & 0 & -X_{\dot{u}} u_r & N_{\dot{r}} r & 0 & -K_{\dot{p}} p \\
-Y_{\dot{v}} v_r & X_{\dot{u}} u_r & 0 & -M_{\dot{q}} q & K_p p & 0
\end{array}\right]
$}
\label{eq_app:C_A}
\end{equation}
The damping matrix $D(\nu_r)$ is modeled as a diagonal matrix, representing proportional damping in each degree of freedom. To account for nonlinear effects at higher speeds, the damping in the surge and sway directions decreases exponentially with the total relative velocity $U_r$ defined in \eqref{eq_app:Total_rel_speed}, introducing velocity-dependent behavior in the damping matrix. The resulting formulation is  
\begin{equation}
\scalebox{0.85}{$
D(\nu_r)=-\operatorname{diag}\Big\{\,X_u e^{-3U_r},\;Y_v e^{-3U_r},\;Z_w,\;K_p,\;M_q,\;N_r\,\Big\}.
$}
\label{eq_app:Damping}
\end{equation}
Each damping coefficient is related to the corresponding element of the inertia matrix parameterized by a characteristic time constant $T_i$, assuming first-order linear damping dynamics:  
\begin{equation}
\begin{aligned}
&X_u=-\frac{m_{11}}{T_1}, \quad Y_v=-\frac{m_{22}}{T_2}, \quad Z_w=-\frac{m_{33}}{T_3}, \\
&K_p=-\frac{m_{44}}{T_4}, \quad M_q=-\frac{m_{55}}{T_5}, \quad N_r=-\frac{m_{66}}{T_6}.
\end{aligned}
\label{eq_app:damp}
\end{equation}

The vector $g(\eta)$ represents the restoring forces and moments acting on the vehicle due to gravity and buoyancy, resulting in nonlinear attitude-dependent behavior. For neutrally buoyant underwater vehicles such as the REMUS~100, these effects are primarily determined by the position of the center of gravity, defined as $r_{bG} = [x_g, y_g, z_g]^T$ and the gravitational force $m g$, relative to the body-fixed origin as in~\eqref{eq_app:submerge}.
\begin{equation}
{g}({\eta})=\left[\begin{array}{c}
0 \\
0 \\
0 \\
z_{\mathrm{g}} mg \cos (\theta) \sin (\phi)-y_{\mathrm{g}} mg \cos (\theta) \cos (\phi) \\
z_{\mathrm{g}} mg \sin (\theta)+x_{\mathrm{g}} mg \cos (\theta) \cos (\phi) \\
x_{\mathrm{g}} mg \cos (\theta) \sin (\phi)-y_{\mathrm{g}} mg \sin (\theta)
\end{array}\right]
\label{eq_app:submerge}
\end{equation}
The restoring forces and moments depend only on roll and pitch angles, as these determine the orientation of the vehicle relative to the vertical gravity and buoyancy forces. The yaw rotation $\psi$ does not influence the restoring effects, since it represents a rotation about the gravity vector itself.

With the control input vector $u=[\delta_s, \delta_r, n_p]^\top$ containing the stern and rudder plane angles, and propeller speed, the generalized forces and moments $\tau$ acting on the AUV can be generated. 
The rudder and stern planes generate drag forces along the surge direction,
\begin{equation}
X_r = -\tfrac{1}{2} \rho U_{rh}^2 A_r C_{L_{\delta_r}} \delta_r^2, \quad
X_s = -\tfrac{1}{2} \rho U_{rv}^2 A_s C_{L_{\delta_s}} \delta_s^2,
\end{equation}
as well as lift forces in sway and heave, respectively,
\begin{equation}
Y_r = -\tfrac{1}{2} \rho U_{rh}^2 A_r C_{L_{\delta_r}} \delta_r, \quad
Z_s = -\tfrac{1}{2} \rho U_{rv}^2 A_s C_{L_{\delta_s}} \delta_s,
\end{equation}
where $\rho$ is the water density, $A_r$ and $A_s$ are the rudder and stern-plane areas, $\delta_r$, $\delta_s$ are their deflection angles, and the relative horizontal and vertical speeds, $U_{rh}$ and $U_{rv}$, are given in \eqref{eq_app:V_H_speed}.  
For positive propeller shaft speeds $n_{p} \ge 0$, the propeller thrust $X_{\mathrm{prop}}$ and torque $K_{\mathrm{prop}}$ can be expressed in simplified form as
\begin{equation}
X_{\mathrm{prop}} = \alpha_X n_{p}^2 + \beta_X n_{p}, 
\quad
K_{\mathrm{prop}} = \alpha_K n_{p}^2 + \beta_K n_{p},
\end{equation}
where the coefficients $\alpha_X$, $\beta_X$, $\alpha_K$, and $\beta_K$ collect the effects of fluid density, propeller and plane geometrics, advance-ratio scheduling, deduction factors, and the thrust/torque coefficients. \\
The resulting generalized force vector is given in~\eqref{eq_app:tau}, where $x_r$, $x_s$ are the longitudinal positions of the rudder and stern planes relative to the body-fixed origin. 
\begin{equation}
\tau = \left[ X_{\mathrm{prop}} + X_r + X_s, 
Y_r , 
Z_s , 
K_{\mathrm{prop}}, 
- x_s Z_s, 
x_r Y_r \right]^{\!\top}
\label{eq_app:tau}
\end{equation}

The lift and drag forces are modeled as functions of the angle of attack $\alpha$, given in~\eqref{eq_app:angle_of_attack}, using the corresponding lift and drag coefficients $C_L(\alpha)$ and $C_D(\alpha)$. The resulting magnitudes are given in~\eqref{eq_app:lift_drag}, where $S$ is the vehicle's reference surface area.
\begin{equation}
F_{\mathrm{lift}} = \tfrac{1}{2} \rho U_{r}^2\,S\,C_L(\alpha)
\qquad 
F_{\mathrm{drag}}  = \tfrac{1}{2} \rho U_{r}^2\,S\,C_D(\alpha)
\label{eq_app:lift_drag}
\end{equation}

The forces are resolved into the body-fixed coordinate system by rotation with the angle of attack $\alpha$, yielding the longitudinal and vertical components
\begin{equation}
\begin{aligned}
\tau_{LD,X} &= -F_{\mathrm{drag}}\cos\alpha \;+\; F_{\mathrm{lift}}\sin\alpha,\\
\tau_{LD,Z} &= -F_{\mathrm{drag}}\sin\alpha \;-\; F_{\mathrm{lift}}\cos\alpha.
\end{aligned}
\label{eq_app:LD_body_components}
\end{equation}
The lift-drag vector $\tau_{LD}$ is then expressed in~\eqref{eq_app:tau_LD} assuming that no side force or rotational moments arise from these components.
\begin{equation}
\tau_{LD}=\left[
\tau_{LD,X}, 0, \tau_{LD,Z}, 0,0,0 
\right]^\top
\label{eq_app:tau_LD}
\end{equation}
The cross-flow drag forces and moments are modeled following the strip-theory formulation by \citep{Faltinsen_1993} and \citep{fossen2021handbook}. The AUV hull is approximated as a series of two-dimensional strips of draft $T = D_{\mathrm{AUV}}$ and length $L = L_{\mathrm{AUV}}$, each experiencing a local flow velocity dependent on the sway and heave components $(v_r,w_r)$ and the corresponding angular rates $(r,q)$. The sectional drag is characterized by the two-dimensional drag coefficient $C_{\mathrm{d}}^{2\mathrm{D}}(v_r,w_r)$, yielding the cross-flow drag integrals in~\eqref{eq_app:CFs}. The total generalized cross-flow drag vector acting on the vehicle is given in~\eqref{eq_app:tau_CF}, representing the hydrodynamic forces and moments in sway, heave, pitch, and yaw due to lateral and vertical flow separation along the hull.
\begin{equation}
\scalebox{0.9}{$
\begin{aligned}
& Y_{CF}=-\tfrac{1}{2} \rho T C_{\mathrm{d}}^{2 \mathrm{D}}(v_r,w_r)\!\int_{-\frac{L}{2}}^{\frac{L}{2}} \!|v_r+x r|\,(v_r+x r)\, \mathrm{d}x\\
& Z_{CF}=-\tfrac{1}{2} \rho T C_{\mathrm{d}}^{2 \mathrm{D}}(v_r,w_r)\!\int_{-\frac{L}{2}}^{\frac{L}{2}} \!|w_r+x q|\,(w_r+x q)\, \mathrm{d}x\\
& M_{CF}=-\tfrac{1}{2} \rho T C_{\mathrm{d}}^{2 \mathrm{D}}(v_r,w_r)\!\int_{-\frac{L}{2}}^{\frac{L}{2}} \!x\,|v_r+x r|\,(v_r+x r)\, \mathrm{d}x\\
& N_{CF}=-\tfrac{1}{2} \rho T C_{\mathrm{d}}^{2 \mathrm{D}}(v_r,w_r)\!\int_{-\frac{L}{2}}^{\frac{L}{2}} \!x\,|w_r+x q|\,(w_r+x q)\, \mathrm{d}x
\end{aligned}
$}
\label{eq_app:CFs}
\end{equation}
\begin{equation}
\tau_{CF}= \left[0, \, Y_{CF}, \, Z_{CF}, \, 0, \, M_{CF}, \, N_{CF}\right]^\top
\label{eq_app:tau_CF}
\end{equation}

\section{Ocean Current model}
\label{app:ocean_current_model}
The current velocity expressed in the body-fixed frame is defined as
\begin{equation}
\nu_c = 
\begin{bmatrix}
u_c ,  v_c ,  w_c ,  p_c ,  q_c ,  r_c
\end{bmatrix}^\top.
\end{equation}
The first three components $(u_c, v_c, w_c)$ represent the linear ocean current velocity, while the remaining components $(p_c, q_c, r_c)$ correspond to rotational current effects. Under the common assumption of irrotational ocean flow, these angular components are negligible and simplify to~\eqref{eq:current_simplified}.
\begin{equation}
\nu_c^b = 
\begin{bmatrix}
u_c ,  v_c ,  w_c
\end{bmatrix}^\top
\label{eq:current_simplified}
\end{equation}
To obtain $\nu_c^b$, the ocean current defined in the NED frame must be transformed using the standard coordinate transformation of~\eqref{eq:Transformation}, as expressed in~\eqref{eq:current_transformation}. The current in the NED frame, ${\nu}_{{c}}^n$, is represented as a superposition of a horizontal component of magnitude $V_c$ and a vertical component $W_c$. The horizontal flow direction is specified by the angle $\beta_{V_c}$, while, due to the separation of components, the vertical current is assumed to act strictly along the $z$-axis, leading to~\eqref{eq:current_simplified_NED}.
\begin{equation}
{\nu}_{{c}}^n=\left[
V_{{c}} \cos \left(\beta_{V_c}\right) \; , \;
V_{{c}} \sin \left(\beta_{V_c}\right) \; , \;
W_{{c}} \right] ^\top
\label{eq:current_simplified_NED}
\end{equation}
\begin{equation}
\nu_c^b=\left(R_b^n\right)^{-1} {\nu}_{{c}}^n
\label{eq:current_transformation}
\end{equation}
Reformulating \eqref{eq:hydrodynamics} into an explicit differential form yields \eqref{eq:ode}, which expresses the dynamics in terms of the time derivatives of the vehicle velocity $\dot{\nu}$ and the ocean current velocity $\dot{\nu}_c$:
\begin{equation}
\begin{aligned}
\dot{\nu}_r &= \dot{\nu} - \dot{\nu}_c \\
&= M^{{\small - }1}\!\big(\tau\!+\!\tau_{LD}\!+\!\tau_{CF}
 {\small - } C(\nu_r)\nu_r {\small - } D(\nu_r)\nu_r {\small - } g(\eta)\big)
\end{aligned}
\label{eq:ode}
\end{equation}
To obtain $\dot{\nu}$, which represents the actual acceleration of the vehicle and is the primary quantity of interest, the derivative of the ocean current velocity $\dot{\nu}_c^b$ must be computed using the skew-symmetric matrix defined in~\eqref{eq_app:Skew-matrix}. Substituting~\eqref{eq:current_derivation} into~\eqref{eq:ode} enables the evaluation of $\dot{\nu}$, completing the formulation of the current-compensated dynamic model.
\begin{equation}
\begin{aligned}
\dot{{\nu}}_{{c}}^b &= -S\!\left(\!\begin{bmatrix}p\\q\\r\end{bmatrix}\!\right)
\begin{bmatrix}
u_c \\ v_c \\ w_c
\end{bmatrix}
=
\begin{bmatrix}
r v_c - q w_c \\[3pt]
-r u_c + p w_c \\[3pt]
q u_c - p v_c
\end{bmatrix}
\end{aligned}
\label{eq:current_derivation}
\end{equation}
\end{document}